\documentclass[preprint,12pt]{elsarticle}

\usepackage{amssymb}
\journal{Phys. Lett. B}

\begin{document}

\begin{frontmatter}



\title{Quark masses from lattice QCD and the study of textures}


\author[GLA]{Craig McNeile}

\address[GLA]{
Bergische Universit\"at Wuppertal, Gaussstr.\,20, D-42119 Wuppertal, Germany
}

\begin{abstract}

I review how the determination 
of quark masses from lattice QCD can be used to study textures
in quark mass matrices. This type of theory relates quark masses
to CKM matrix elements. I demonstrate how the recent
precision results from the HPQCD and MILC 
collaborations for quark masses
can be used to test some of these ideas.

\end{abstract}


\end{frontmatter}


\section{Introduction}
\label{se:intro}

The existence of different families of quarks and leptons
is a puzzle. It is important to try to explain why
the CKM matrix is diagonally dominant and 
to understand
the values of the masses of the quarks.
One way to understand 
the additional flavours is to look for some symmetry between
the 
families~\cite{Froggatt:1978nt,Binetruy:1994ru,Antusch:2009hq,Ramond:1993kv}.
In the quark sector the way that the symmetries
are searched for is to look for connections between
the quark masses and the CKM matrix elements.
There has been a long history of looking for
patterns in the quark masses and the CKM matrix
elements~\cite{Fritzsch:1979zq,Wilczek:1977uh}.

Given that there is no theory that predicts symmetries
between families, the subject is driven by the size of
the errors on the CKM matrix elements and quark masses.
The reduced errors bars on the CKM matrix from the
experimental results from B factories~\cite{Charles:2004jd}
and improved
theoretical predictions from techniques such as lattice
QCD have already ruled out some proposed relationships 
between quark masses and CKM matrix elements.
The PDG quote the current error on the strange quark mass as
around 30\%. However, 
this is accurate enough to rule out one of the original
predictions:
$V_{cb} \sim \frac{m_s}{m_b}$
of a 6 texture model~\cite{Fritzsch:1977vd}.
The error on the strange quark mass
should be compared to the largest error on a CKM matrix element or ratio
of CKM matrix, used in this paper, of 10\%.
The subject of mass and flavor mixing has been reviewed
by Fritzsch and Xing~\cite{Fritzsch:1999ee}, 
Froggatt~\cite{Froggatt:2003ef}, and 
Babu~\cite{Babu:2009fd}.

Recently the HPQCD collaborations have determined the 
masses of the strange, charm, and bottom
quarks with an error of under 2\% from unquenched lattice
QCD~\cite{McNeile:2010ji,Davies:2009ih,Allison:2008xk}. 
The MILC collaboration had previously determined the 
ratio and sum of the masses of the up and down quarks~\cite{Aubin:2004fs}.
The unprecedented precision stems from 
fitting lattice correlators in the continuum limit
to continuum perturbation theory that depends on the masses of the charm and
bottom quarks. The powerful techniques of multiloop QCD
in the continuum have been used to compute the correlators
to 4 loop order, hence this reduces one of the major systematic
errors in lattice QCD calculations of quark masses, the conversion
from the lattice results to the $\overline{MS}$ scheme. 
The light and strange
quark masses are determined using the continuum limit of 
the ratios of quark masses. The suggestion to use the 
ratio of the light quark masses to the charm mass was
originally suggested in the famous review article by
Gasser and Leutwyler~\cite{Gasser:1982ap}. The basis
of these results are generation of gauge configurations
with 2+1 flavours of sea quark with light pion masses and
multiple lattice spacings
by the MILC 
collaboration~\cite{Bernard:2001av,Aubin:2004wf,Bazavov:2009bb}. 
Particularly important are the new gauge configurations 
at lattice spacings of 0.06 and 0.045 fm,
generated by the MILC collaboration~\cite{Bazavov:2009bb}, 
that were
crucial to taking a continuum limit for 
calculations with valence charm and bottom quarks.

The lattice QCD calculations 
by the HPQCD collaboration
have been tested by prediction
of the mass of the $B_c$ meson~\cite{Allison:2004be}
and the $\eta_b$ meson~\cite{Gray:2005ur}. A summary
of the mass spectrum of heavy-heavy and heavy-light 
mesons is in~\cite{Gregory:2009hq}. The decay constants
of the pion and kaon have been accurately computed~\cite{Follana:2007uv}.
The value of $\alpha_s$ extracted 
from the  HPQCD collaboration~\cite{Davies:2008sw,McNeile:2010ji}
is consistent with other non-lattice determinations in~\cite{Bethke:2009jm}.
The staggered fermion formalism potentially
has a problem with the technical issue of ``rooting of determinants''.
However, no theoretical work has found any 
problems~\cite{Kronfeld:2007ek,Sharpe:2006re}.
So for the arguments given in this paragraph the use
of HPQCD's quark masses is reasonable,
rather than using the more conservative errors
on the quark masses quoted by the PDG.
Scholz~\cite{Scholz:2009yz} 
and Leutwyler~\cite{Leutwyler:2009jg}
have recently reviewed 
the status of lattice QCD calculations of quark masses.

Apart from one simple example in section~\ref{se:conc},
I don't discuss the relations between the quark masses
and lepton masses predicted by some grand unified
theories~\cite{Georgi:1979df}. This would be an important,
but separate study and also require running the masses
to the GUT scale~\cite{Fusaoka:1998vc} in a model dependent
way. See~\cite{Antusch:2009gu,Ross:2007az}
for two recent studies.

In this paper I use the new precision results
for quark masses from the 
HPQCD and MILC 
collaborations~\cite{Allison:2008xk,Davies:2009ih,Aubin:2004fs}
to test some of the relations between CKM
matrix elements and quark masses 
proposed by Chkareuli and Froggatt~~\cite{Chkareuli:1998sa},
and
Fritzsch and Xing~\cite{Fritzsch:2002ga}.
A goal
of this project was to find out how accurately do we need
to know the masses of the up, down and strange quarks.

\section{An introduction to textures} \label{se:texture}

The part of the standard model Lagrangian that describes
the quark masses is
\begin{equation}
{\cal L} = - \overline{u}^i_L (M_u)_{ij}  u^j_R
- \overline{d}^i_L (M_d)_{ij}  d^j_R
+ \mbox{h. c.}
\end{equation}
where $j$ is a index over flavour.

The mass matrices $M_u$ and $M_d$ are diagonalized
to obtain the quark masses,
\begin{eqnarray}
V_{uL} M_u V_{uR}^\dagger & = & diag( m_u, m_c, m_t) \\
V_{dL} M_d V_{dR}^\dagger & = & diag( m_d, m_s, m_b)
\end{eqnarray}
by the order 3 unitary matrices $V_{uL}$, $V_{uR}$,
$V_{dL}$, and $V_{dR}$.

The CKM matrix is
\begin{equation}
U_{CKM} = V_{uL} V_{dL}^\dagger
\end{equation}
A key prediction of the standard model is
that the $U_{CKM}$ matrix is unitary.
The experimental program at the B factories have
not found any significant deviation from
unitarity~\cite{Charles:2004jd}.
Although there are perhaps some 
hints~\cite{VandeWater:2009uc,Laiho:2009eu}.
The CKM matrix can be determined using the 
results from
lattice QCD~\cite{VandeWater:2009uc}.

The idea is to look for some special structure in the
mass mixing matrices $M_u$ and $M_d$.  
For example there
could be zeros in the mass matrices, 
which are known as textures~\cite{Fritzsch:1999ee}.
One concern about this is
that a special basis is chosen and the patterns
could be removed by a 
transformation~\cite{Branco:1999nb,Jarlskog:2006za}. 
The hope is that
in some basis the physics is more transparent.

\begin{table}[tb]
\centering
\begin{tabular}{|c|c|c|} \hline
Quantity & HPQCD/MILC  &  PDG \\
\hline
$m_u(\mbox{2 GeV})$ &  $2.01  \pm 0.10$     &  $2.55 \pm 1.05$   \\
$m_d(\mbox{2 GeV})$ &  $4.77 \pm 0.15$      &  $5.04 \pm 1.54$   \\
$m_s(\mbox{2 GeV})$ &  $92.2 \pm 1.3 $      &  $105  \pm 35$     \\
$m_c(m_c)$ GeV      &  $1.273 \pm 0.006 $   &  $1.27 \pm 0.11$   \\
$m_b(m_b)$ GeV      &  $4.164 \pm 0.023  $  &  $4.20  \pm 0.17$  \\
$m_t(m_t)$ GeV      &  -                    &  $160 \pm 3 $      \\
$m_c/m_s$           &  $11.85 \pm 0.16$     &  -                 \\
$m_u/m_d$           &  $0.42 \pm 0.04$      & 0.35 - 0.6         \\
$m_s/m_l$           &  $27.2 \pm 0.03$      & 25 - 30            \\
$m_b/m_c$           &  $4.51 \pm 0.04$      & -            \\
\hline
\end{tabular}
\caption{Summary of quark masses and ratios of quark masses from the 
HPQCD and MILC 
collaborations~\cite{McNeile:2010ji,Davies:2009ih,Allison:2008xk,Aubin:2004fs},
the PDG~\cite{Amsler:2008zzb}, and Langenfeld et al.~\cite{Langenfeld:2009wd}}
\label{tb:quarkMass}
\end{table}

In table~\ref{tb:quarkMass}, I list the quark masses used in this study.
I use the quark masses from the HPQCD and MILC collaborations
and the ranges from the PDG~\cite{Amsler:2008zzb}. The $m_l$ mass
is the average of quark masses of the up and down quarks.
I use the value for the top quark mass,
in the $\overline{MS}$ scheme,
from
the work of Langenfeld et al.~\cite{Langenfeld:2009wd}.
Many of the results from the HPQCD and MILC collaborations
are ratio of quark masses, so where possible I use
products of ratios of quark masses to compute the appropriate
combination of quark masses.
Otherwise I use 
the RunDec package~\cite{Chetyrkin:2000yt}
(using perturbative results from
from~\cite{vanRitbergen:1997va,Czakon:2004bu})
to run the quark masses to 2 GeV. 
I use the value of $\alpha_s^{\overline{MS}}(M_Z,n_f=5)$ 
= 0.1184(6)
from the HPQCD collaboration~\cite{Davies:2008sw}.

\begin{table}[tb]
\centering
\begin{tabular}{|c|c|c|} \hline
$V_{us}$ = $0.2255 \pm 0.0019$ &
$V_{ub}$ = $(3.93 \pm 0.36) 10^{-3} $ &
$V_{cb}$ = $(41.2 \pm 1.1) 10^{-3} $
\\
\hline
\end{tabular}
\caption{Values of the CKM matrix taken from the PDG~\cite{Amsler:2008zzb}}
\label{tb:ckm}
\end{table}

The CKM matrix are related to Yukawa couplings so must
be renormalised~\cite{Almasy:2008ep}.
I don't include any running of the 
CKM matrix~\cite{Balzereit:1998id}, because the effects are
thought to be small at low 
energies~\cite{Fritzsch:1999ee}.

\section{Results}
\label{se:froggatt}

In this section I will investigate the model
for CKM parameters and quark masses 
introduced by Chkareuli and Froggatt~\cite{Chkareuli:1998sa}.
They proposed the following ansatz for the up and down mass matrices.
\begin{eqnarray}
M_u & = & \left(
\begin{array}{ccc} 
0   & 0         & \sqrt{m_u m_t} e^{ i c_U }  \\ 
0   & m_c       & 0 \\ 
\sqrt{m_u m_t} e^{ -i c_U } & 0  & m_t - m_u
\end{array} 
\right)
\label{eq:Massup}
\\
M_d & = & \left(
\begin{array}{ccc} 
0 & \sqrt{m_d m_s}e^{ i a_D } & 0 \\ 
\sqrt{m_d m_s}e^{ -i a_D } & m_s & \sqrt{m_d m_b}e^{ i b_D } \\ 
0 & \sqrt{m_d m_b}e^{ -i b_D } & m_b - m_d
\end{array} 
\right)
\label{eq:Massdown}
\end{eqnarray}
where $c_U$, $a_D$, $b_D$ are phases.
The ansatz is based on an idea called Lightest
Flavor Mixing. See the original paper for the motivation
for the mass matrices~\cite{Chkareuli:1998sa}.

Their model predicts the following relationships between
the quark masses and CKM matrix elements.
\begin{eqnarray}
V_{us} & \sim & \sqrt{\frac{m_d}{m_s} } =  \sqrt{\frac{2}{1 +
    \frac{m_u}{m_d}} \frac{m_l}{m_s} }  \label{eq:frogpredictus} \\
V_{cb} & \sim & \sqrt{\frac{m_d}{m_b} } =
\sqrt{\frac{m_d}{m_s}\frac{m_s}{m_c}\frac{m_c}{m_b}
}  \label{eq:frogpredictcb} \\
V_{ub} & \sim & \sqrt{\frac{m_u}{m_t} } \label{eq:frogpredictub}
\end{eqnarray}

 \begin{figure}
\begin{center}
\includegraphics[scale=0.4,angle=270]{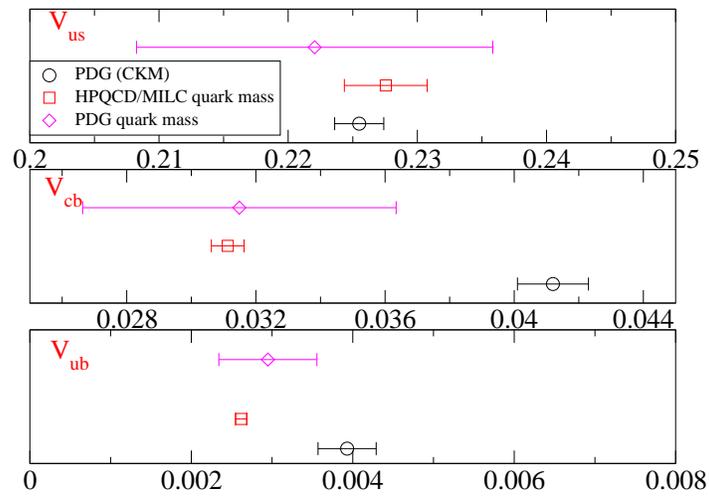}
\end{center}
 \caption {
Test of the 
relationships~\ref{eq:frogpredictus},~\ref{eq:frogpredictcb}, and
~\ref{eq:frogpredictub},
between masses and CKM 
matrix elements predicted by Chkareuli and Froggatt~\cite{Chkareuli:1998sa}.
The circles are the results for the CKM matrix elements. The squares
and diamonds are the predictions for the CKM matrix elements in terms
of quarks masses from the HPQCD/MILC collaborations 
and the PDG respectively.}
\label{fig:Frogg}
 \end{figure}

In figure~\ref{fig:Frogg}, I plot tests of the relationships
in equations~\ref{eq:frogpredictus},~\ref{eq:frogpredictcb}
 and~\ref{eq:frogpredictub}
of Chkareuli and Froggatt~\cite{Chkareuli:1998sa}.
The plots shows that the effect of the reduced
errors on the quark masses from the HPQCD and MILC collaborations,
over the PDG values.
Figure~\ref{fig:Frogg} shows that the prediction
for $V_{cb}$ disagrees with the result from quark mass
prediction of the HPQCD collaboration at the 10$\sigma$ level.

Chkareuli and Froggatt~\cite{Chkareuli:1998sa}
have another ansatz for the quark mass matrices, 
that predicts the relationship in 
equation~\ref{eq:frogpredictI}.
%%%
%%%
\begin{equation}
\frac{\mid V_{ub} \mid}{\mid V_{cb} \mid}
  \sim  \sqrt{\frac{m_u}{m_c} } 
 \label{eq:frogpredictI}
\end{equation}
The quark masses from the PDG give
$\sqrt{\frac{m_u}{m_c}} =$ 0.049(10),
this disagrees with 
$\frac{\mid V_{ub} \mid}{\mid V_{cb} \mid}$=0.095(9)
at the $4\sigma$ level.

\begin{table}[tb]
\centering
\begin{tabular}{|c|c||c|c|} \hline
\multicolumn{2}{|c|}{CKM} &
\multicolumn{2}{|c|}{Quark mass} \\
Element & Value &  PDG  & HPQCD/MILC \\ \hline
$\frac{\mid V_{ub} \mid}{\mid V_{cb} \mid}$ & 0.095(9) & 
0.069(15) & 0.061(2) \\ \hline
%%%
$\mid V_{us} \mid$ & 0.2255(19)  & 0.22(4) & 0.232(3)  \\ \hline
$\frac{\mid V_{td} \mid}{\mid V_{ts} \mid}$  & 0.209(6) & 
 0.22(1) & 0.228(3) 
\\ \hline
\end{tabular}
\caption{Comparison of CKM matrix elements from experiment
with estimates from
equation~\ref{eq:VubCB},~\ref{eq:Vus} and ~\ref{eq:VtdVts},
using the quark masses from the PDG and those from the HPQCD
and MILC collaborations.}
\label{tb:impPRED}
\end{table}

Fritzsch and Xing~\cite{Fritzsch:2002ga} investigated a 
number of 4 texture mass matrices. A subset of the
relations between quark masses and CKM matrix elements,
that they derived, are below.
\begin{eqnarray}
\frac{\mid V_{ub} \mid}{\mid V_{cb} \mid} & \sim &
\sqrt{2 \frac{m_u}{m_c} } \label{eq:VubCB} \\
\mid V_{us} \mid & \sim & 
\sqrt{ m_u / m_c + m_d/m_s } \label{eq:Vus} \\
\frac{\mid V_{td} \mid}{\mid V_{ts} \mid} & \sim &
\sqrt{\frac{m_d}{m_s} } \label{eq:VtdVts}
\end{eqnarray}

The numerical comparison of the
relations between CKM matrix elements and
quark masses derived by Fritzsch and Xing~\cite{Fritzsch:2002ga}
are in table~\ref{tb:impPRED}.
The improved prediction for 
$\frac{\mid V_{ub} \mid}{\mid V_{cb} \mid}$
in equation ~\ref{eq:VubCB} now only disagrees
with the prediction of the quark masses from
the PDG at the $1.6\sigma$ level. 
However, using the 
more accurate quark masses from the HPQCD/MILC collaboration,
the predictions of Fritzsch and Xing~\cite{Fritzsch:2002ga}
for $\frac{\mid V_{ub} \mid}{\mid V_{cb} \mid}$
and
$\frac{\mid V_{td} \mid}{\mid V_{ts} \mid}$
disagree with experiment by over 3$\sigma$.

\section{Conclusions}
\label{se:conc}

In this paper I have used the new high precision results for
quark masses from the HPQCD and MILC collaborations to test some
predictions for relations between CKM matrix elements.
To illustrate the point I used some 
older ansatze~\cite{Chkareuli:1998sa,Fritzsch:2002ga}
 for the quark matrices. The new accurate results
for the quark masses from HPQCD and MILC essentially
ruled out the two models.
Many other studies of textures, 
such as~\cite{Antusch:2009hq,EmmanuelCosta:2009bx,Couture:2009it}, 
would benefit
from more accurate values for the quark masses.

At the moment the smallest errors on the quark masses come from one 
unquenched lattice QCD calculation. 
Although in the introduction I briefly reviewed
how well validated HPQCD's calculation were against experiment,
it would clearly be preferable to have accurate 
results from other lattice formalisms. Recent reviews
of dynamical lattice QCD calculations show that many groups
have access to data with pion masses at least at 300 MeV,
multiple lattice spacings and 
volumes~\cite{Jansen:2008vs,Jung:2010jt}, so the prospects
for reductions in the errors of quark masses from other
formulations is good.

It would be interesting to also use precision lattice QCD
results for quark masses to test the predictions
of grand unified theories. 
In GUTs there could be relationships between lepton and
quark masses at some renormalization scale.
For example Georgi and Jarlskog 
suggested~\cite{Georgi:1979df} (see Babu~\cite{Babu:2009fd} for
a review),
a relationship between the masses of the down and strange quarks,
and the masses of the muon and electron masses.
\begin{equation}
\frac{m_s}{m_d} = \frac{1}{9} \frac{m_\mu}{m_e}
\label{eq:GLrelats}
\end{equation}
The factor of $\frac{1}{9}$ in equation~\ref{eq:GLrelats}
is from the square of the number of colours.
Experimentally, $\frac{1}{9} \frac{m_\mu}{m_e}$ = 21.3.
The new lattice QCD numbers from the HPQCD collaboration,
give $\frac{m_s}{m_d}$ = 19.3(5), while the current ranges
from the PDG give $\frac{m_s}{m_d}$ = 20.3(2.5). So the
new results from HPQCD for the quark masses are inconsistent
with the relation in equation~\ref{eq:GLrelats}
 at the 4$\sigma$ level at low energy.
However, equation~\ref{eq:GLrelats} only needs to hold
at the unification scale and a more detailed study would
include the effects of additional particles, such as
those from SUSY 
models~\cite{Antusch:2009gu,Ross:2007az}.

To test the texture relations the errors on the ratios
of quark masses need to be at least the relative size of
the errors on the relevant CKM matrix elements.  For example
the $V_{us}$ CKM matrix element is currently known to under 1\% accuracy,
so a similar accuracy is required for ratios of quark masses
for a good test of the predictions of textures. The results from improved
lattice QCD calculations will also reduce the errors
on CKM matrix elements, such as $V_{us}$~\cite{Lubicz:2010nx}.

The predictions in equations~\ref{eq:frogpredictus},
~\ref{eq:frogpredictcb}, ~\ref{eq:frogpredictub}
and~\ref{eq:GLrelats}
critically depend
on the values of the masses of the up and down quarks. 
Currently the lattice results for the
masses of the up and down quarks are based on the method
developed by the MILC collaborations~\cite{Aubin:2004fs}.
In particular the majority of the error on 
the ratio $m_u/m_d$ is due their treatment of electromagnetic
effects.
There are new lattice QCD 
calculations~\cite{Basak:2008na,Zhou:2009ku}
that explicitly include
electromagnetism ~\cite{Duncan:1996xy,Blum:2007cy,Basak:2008na},
or estimate the corrections~\cite{Shintani:2008qe,Boyle:2009xi}
that should reduce the errors on
the masses of the up and down quarks obtained from
lattice QCD calculations.

As reviewed by Babu~\cite{Babu:2009fd}, the structure
of the quark mass matrices could be caused by 
``Flavon fields''~\cite{Froggatt:1978nt}.
Although the dynamics that generates the quark mass
matrices could be
at the Planck energy, there could also be measurable effects
at the LHC~\cite{Giudice:2008uua,Raidal:2008jk,Babu:1999me}.

\section{Acknowledgments}

I thank Christine Davies for valuable comments on this paper.
This research used resources of the Argonne Leadership Computing
Facility at Argonne National Laboratory, which is supported by the
Office of Science of the U.S. Department of Energy under contract
DE-AC02-06CH11357.  

%%\bibliographystyle{h-physrev2}
%\bibliography{texture} 

\begin{thebibliography}{10}

\bibitem{Froggatt:1978nt}
C.~D. Froggatt and H.~B. Nielsen,
\newblock Nucl. Phys. {\bf B147}, 277 (1979),
\newblock %%CITATION = NUPHA,B147,277;%%.

\bibitem{Binetruy:1994ru}
P.~Binetruy and P.~Ramond,
\newblock Phys. Lett. {\bf B350}, 49 (1995), hep-ph/9412385,
\newblock %%CITATION = HEP-PH/9412385;%%.

\bibitem{Antusch:2009hq}
S.~Antusch, S.~F. King, M.~Malinsky, and M.~Spinrath,
\newblock Phys. Rev. {\bf D81}, 033008 (2010), 0910.5127,
\newblock %%CITATION = 0910.5127;%%.

\bibitem{Ramond:1993kv}
P.~Ramond, R.~G. Roberts, and G.~G. Ross,
\newblock Nucl. Phys. {\bf B406}, 19 (1993), hep-ph/9303320,
\newblock %%CITATION = HEP-PH/9303320;%%.

\bibitem{Fritzsch:1979zq}
H.~Fritzsch,
\newblock Nucl. Phys. {\bf B155}, 189 (1979),
\newblock %%CITATION = NUPHA,B155,189;%%.

\bibitem{Wilczek:1977uh}
F.~Wilczek and A.~Zee,
\newblock Phys. Lett. {\bf B70}, 418 (1977),
\newblock %%CITATION = PHLTA,B70,418;%%.

\bibitem{Charles:2004jd}
CKMfitter Group, J.~Charles {\em et~al.},
\newblock Eur. Phys. J. {\bf C41}, 1 (2005), hep-ph/0406184,
\newblock %%CITATION = HEP-PH/0406184;%%.

\bibitem{Fritzsch:1977vd}
H.~Fritzsch,
\newblock Phys. Lett. {\bf B73}, 317 (1978),
\newblock %%CITATION = PHLTA,B73,317;%%.

\bibitem{Fritzsch:1999ee}
H.~Fritzsch and Z.-z. Xing,
\newblock Prog. Part. Nucl. Phys. {\bf 45}, 1 (2000), hep-ph/9912358,
\newblock %%CITATION = HEP-PH/9912358;%%.

\bibitem{Froggatt:2003ef}
C.~D. Froggatt,
\newblock Surveys High Energ. Phys. {\bf 18}, 77 (2003), hep-ph/0307138,
\newblock %%CITATION = HEP-PH/0307138;%%.

\bibitem{Babu:2009fd}
K.~S. Babu,
\newblock (2009), 0910.2948,
\newblock %%CITATION = 0910.2948;%%.

\bibitem{McNeile:2010ji}
C.~McNeile, C.~T.~H. Davies, E.~Follana, K.~Hornbostel, and G.~P. Lepage,
\newblock (2010), 1004.4285,
\newblock %%CITATION = 1004.4285;%%.

\bibitem{Davies:2009ih}
C.~T.~H. Davies {\em et~al.},
\newblock Phys. Rev. Lett. {\bf 104}, 132003 (2010), 0910.3102,
\newblock %%CITATION = 0910.3102;%%.

\bibitem{Allison:2008xk}
HPQCD, I.~Allison {\em et~al.},
\newblock Phys. Rev. {\bf D78}, 054513 (2008), 0805.2999,
\newblock %%CITATION = 0805.2999;%%.

\bibitem{Aubin:2004fs}
MILC, C.~Aubin {\em et~al.},
\newblock Phys. Rev. {\bf D70}, 114501 (2004), hep-lat/0407028,
\newblock %%CITATION = HEP-LAT/0407028;%%.

\bibitem{Gasser:1982ap}
J.~Gasser and H.~Leutwyler,
\newblock Phys. Rept. {\bf 87}, 77 (1982),
\newblock %%CITATION = PRPLC,87,77;%%.

\bibitem{Bernard:2001av}
C.~W. Bernard {\em et~al.},
\newblock Phys. Rev. {\bf D64}, 054506 (2001), hep-lat/0104002,
\newblock %%CITATION = HEP-LAT/0104002;%%.

\bibitem{Aubin:2004wf}
C.~Aubin {\em et~al.},
\newblock Phys. Rev. {\bf D70}, 094505 (2004), hep-lat/0402030,
\newblock %%CITATION = HEP-LAT/0402030;%%.

\bibitem{Bazavov:2009bb}
A.~Bazavov {\em et~al.},
\newblock (2009), 0903.3598,
\newblock %%CITATION = 0903.3598;%%.

\bibitem{Allison:2004be}
HPQCD, I.~F. Allison {\em et~al.},
\newblock Phys. Rev. Lett. {\bf 94}, 172001 (2005), hep-lat/0411027,
\newblock %%CITATION = HEP-LAT/0411027;%%.

\bibitem{Gray:2005ur}
A.~Gray {\em et~al.},
\newblock Phys. Rev. {\bf D72}, 094507 (2005), hep-lat/0507013,
\newblock %%CITATION = HEP-LAT/0507013;%%.

\bibitem{Gregory:2009hq}
E.~B. Gregory {\em et~al.},
\newblock Phys. Rev. Lett. {\bf 104}, 022001 (2010), 0909.4462,
\newblock %%CITATION = 0909.4462;%%.

\bibitem{Follana:2007uv}
HPQCD, E.~Follana, C.~T.~H. Davies, G.~P. Lepage, and J.~Shigemitsu,
\newblock Phys. Rev. Lett. {\bf 100}, 062002 (2008), 0706.1726,
\newblock %%CITATION = 0706.1726;%%.

\bibitem{Davies:2008sw}
HPQCD, C.~T.~H. Davies {\em et~al.},
\newblock Phys. Rev. {\bf D78}, 114507 (2008), 0807.1687,
\newblock %%CITATION = 0807.1687;%%.

\bibitem{Bethke:2009jm}
S.~Bethke,
\newblock Eur. Phys. J. {\bf C64}, 689 (2009), 0908.1135,
\newblock %%CITATION = 0908.1135;%%.

\bibitem{Kronfeld:2007ek}
A.~S. Kronfeld,
\newblock PoS {\bf LAT2007}, 016 (2007), 0711.0699,
\newblock %%CITATION = 0711.0699;%%.

\bibitem{Sharpe:2006re}
S.~R. Sharpe,
\newblock PoS {\bf LAT2006}, 022 (2006), hep-lat/0610094,
\newblock %%CITATION = HEP-LAT/0610094;%%.

\bibitem{Scholz:2009yz}
E.~E. Scholz,
\newblock (2009), 0911.2191,
\newblock %%CITATION = 0911.2191;%%.

\bibitem{Leutwyler:2009jg}
H.~Leutwyler,
\newblock (2009), 0911.1416,
\newblock %%CITATION = 0911.1416;%%.

\bibitem{Georgi:1979df}
H.~Georgi and C.~Jarlskog,
\newblock Phys. Lett. {\bf B86}, 297 (1979),
\newblock %%CITATION = PHLTA,B86,297;%%.

\bibitem{Fusaoka:1998vc}
H.~Fusaoka and Y.~Koide,
\newblock Phys. Rev. {\bf D57}, 3986 (1998), hep-ph/9712201,
\newblock %%CITATION = HEP-PH/9712201;%%.

\bibitem{Antusch:2009gu}
S.~Antusch and M.~Spinrath,
\newblock Phys. Rev. {\bf D79}, 095004 (2009), 0902.4644,
\newblock %%CITATION = 0902.4644;%%.

\bibitem{Ross:2007az}
G.~Ross and M.~Serna,
\newblock Phys. Lett. {\bf B664}, 97 (2008), 0704.1248,
\newblock %%CITATION = 0704.1248;%%.

\bibitem{Chkareuli:1998sa}
J.~L. Chkareuli and C.~D. Froggatt,
\newblock Phys. Lett. {\bf B450}, 158 (1999), hep-ph/9812499,
\newblock %%CITATION = HEP-PH/9812499;%%.

\bibitem{Fritzsch:2002ga}
H.~Fritzsch and Z.-z. Xing,
\newblock Phys. Lett. {\bf B555}, 63 (2003), hep-ph/0212195,
\newblock %%CITATION = HEP-PH/0212195;%%.

\bibitem{VandeWater:2009uc}
R.~S. Van~de Water,
\newblock (2009), 0911.3127,
\newblock %%CITATION = 0911.3127;%%.

\bibitem{Laiho:2009eu}
J.~Laiho, E.~Lunghi, and R.~S. Van~de Water,
\newblock Phys. Rev. {\bf D81}, 034503 (2010), 0910.2928,
\newblock %%CITATION = 0910.2928;%%.

\bibitem{Branco:1999nb}
G.~C. Branco, D.~Emmanuel-Costa, and R.~Gonzalez~Felipe,
\newblock Phys. Lett. {\bf B477}, 147 (2000), hep-ph/9911418,
\newblock %%CITATION = HEP-PH/9911418;%%.

\bibitem{Jarlskog:2006za}
C.~Jarlskog,
\newblock Phys. Scripta {\bf T127}, 64 (2006), hep-ph/0606050,
\newblock %%CITATION = HEP-PH/0606050;%%.

\bibitem{Amsler:2008zzb}
Particle Data Group, C.~Amsler {\em et~al.},
\newblock Phys. Lett. {\bf B667}, 1 (2008),
\newblock %%CITATION = PHLTA,B667,1;%%.

\bibitem{Langenfeld:2009wd}
U.~Langenfeld, S.~Moch, and P.~Uwer,
\newblock Phys. Rev. {\bf D80}, 054009 (2009), 0906.5273,
\newblock %%CITATION = 0906.5273;%%.

\bibitem{Chetyrkin:2000yt}
K.~G. Chetyrkin, J.~H. Kuhn, and M.~Steinhauser,
\newblock Comput. Phys. Commun. {\bf 133}, 43 (2000), hep-ph/0004189,
\newblock %%CITATION = HEP-PH/0004189;%%.

\bibitem{vanRitbergen:1997va}
T.~van Ritbergen, J.~A.~M. Vermaseren, and S.~A. Larin,
\newblock Phys. Lett. {\bf B400}, 379 (1997), hep-ph/9701390,
\newblock %%CITATION = HEP-PH/9701390;%%.

\bibitem{Czakon:2004bu}
M.~Czakon,
\newblock Nucl. Phys. {\bf B710}, 485 (2005), hep-ph/0411261,
\newblock %%CITATION = HEP-PH/0411261;%%.

\bibitem{Almasy:2008ep}
A.~A. Almasy, B.~A. Kniehl, and A.~Sirlin,
\newblock Phys. Rev. {\bf D79}, 076007 (2009), 0811.0355,
\newblock %%CITATION = 0811.0355;%%.

\bibitem{Balzereit:1998id}
C.~Balzereit, T.~Mannel, and B.~Plumper,
\newblock Eur. Phys. J. {\bf C9}, 197 (1999), hep-ph/9810350,
\newblock %%CITATION = HEP-PH/9810350;%%.

\bibitem{EmmanuelCosta:2009bx}
D.~Emmanuel-Costa and C.~Simoes,
\newblock Phys. Rev. {\bf D79}, 073006 (2009), 0903.0564,
\newblock %%CITATION = 0903.0564;%%.

\bibitem{Couture:2009it}
G.~Couture, C.~Hamzaoui, S.~S.~Y. Lu, and M.~Toharia,
\newblock Phys. Rev. {\bf D81}, 033010 (2010), 0910.3132,
\newblock %%CITATION = 0910.3132;%%.

\bibitem{Jansen:2008vs}
K.~Jansen,
\newblock PoS {\bf LATTICE2008}, 010 (2008), 0810.5634,
\newblock %%CITATION = 0810.5634;%%.

\bibitem{Jung:2010jt}
C.~Jung,
\newblock (2010), 1001.0941,
\newblock %%CITATION = 1001.0941;%%.

\bibitem{Lubicz:2010nx}
V.~Lubicz,
\newblock PoS {\bf LAT2009}, 013 (2009), 1004.3473,
\newblock %%CITATION = 1004.3473;%%.

\bibitem{Basak:2008na}
MILC, S.~Basak {\em et~al.},
\newblock PoS {\bf LATTICE2008}, 127 (2008), 0812.4486,
\newblock %%CITATION = 0812.4486;%%.

\bibitem{Zhou:2009ku}
R.~Zhou and S.~Uno,
\newblock PoS {\bf LAT2009}, 182 (2009), 0911.1541,
\newblock %%CITATION = 0911.1541;%%.

\bibitem{Duncan:1996xy}
A.~Duncan, E.~Eichten, and H.~Thacker,
\newblock Phys. Rev. Lett. {\bf 76}, 3894 (1996), hep-lat/9602005,
\newblock %%CITATION = HEP-LAT/9602005;%%.

\bibitem{Blum:2007cy}
T.~Blum, T.~Doi, M.~Hayakawa, T.~Izubuchi, and N.~Yamada,
\newblock Phys. Rev. {\bf D76}, 114508 (2007), 0708.0484,
\newblock %%CITATION = 0708.0484;%%.

\bibitem{Shintani:2008qe}
JLQCD, E.~Shintani {\em et~al.},
\newblock Phys. Rev. Lett. {\bf 101}, 242001 (2008), 0806.4222,
\newblock %%CITATION = 0806.4222;%%.

\bibitem{Boyle:2009xi}
RBC, P.~A. Boyle, L.~Del~Debbio, J.~Wennekers, and J.~M. Zanotti,
\newblock Phys. Rev. {\bf D81}, 014504 (2010), 0909.4931,
\newblock %%CITATION = 0909.4931;%%.

\bibitem{Giudice:2008uua}
G.~F. Giudice and O.~Lebedev,
\newblock Phys. Lett. {\bf B665}, 79 (2008), 0804.1753,
\newblock %%CITATION = 0804.1753;%%.

\bibitem{Raidal:2008jk}
M.~Raidal {\em et~al.},
\newblock Eur. Phys. J. {\bf C57}, 13 (2008), 0801.1826,
\newblock %%CITATION = 0801.1826;%%.

\bibitem{Babu:1999me}
K.~S. Babu and S.~Nandi,
\newblock Phys. Rev. {\bf D62}, 033002 (2000), hep-ph/9907213,
\newblock %%CITATION = HEP-PH/9907213;%%.

\end{thebibliography}

\end{document}